\documentclass[preprint,eqsecnum,showpacs,aps]{revtex4}   % shows pacs
\usepackage{amsmath}
\def\beq{\begin{equation}}
\def\eeq{\end{equation}}
\def\bea{\begin{eqnarray}}
\def\eea{\end{eqnarray}}
\def\nnu{\nonumber}
\def\tst{\textstyle}
\def\dst{\displaystyle}

\def\olc{\onlinecite}

\def\eno#1{Eq.~(\ref{#1})}
\def\Eno#1{Equation (\ref{#1})}
\def\etwo#1#2{Eqs.~(\ref{#1}) and (\ref{#2})}

\def\Sno#1{Sec.~\ref{#1}}

\def\gtwid{\mathrel{\raise.3ex\hbox{$>$\kern75em\lower1ex\hbox{$\sim$}}}}
\def\ltwid{\mathrel{\raise.3ex\hbox{$<$\kern-.75em\lower1ex\hbox{$\sim$}}}}

\def\al{\alpha}

\def\gam{\gamma}
\def\dta{\delta}
\def\eps{\epsilon}

\def\tta{\theta}

\def\vph{\varphi}

\def\Gam{\Gamma}
\def\Dta{\Delta}

\def\apx{\approx}
\def\ptl{\partial}
\def\hf{\frac{1}{2}}
\def\tshf{{\tst\hf}}

\def\tofro{\leftrightarrow}

\def\grad{\nabla}

\def\part#1#2{\frac{\ptl#1}{\ptl#2}}

\def\adag{a^{\dagger}}
\def\ham{{\mathcal{H}}}
\def\ket#1{|#1\rangle}

\def\bra#1{\langle#1|}

\def\olap#1#2{\langle#1|#2\rangle}
\def\kb#1#2{\ket{#1}\bra{#2}}

\def\mel#1#2#3{\langle#1|#2|#3\rangle}

\def\Tr{{\rm Tr}\,}

\def\bJ{{\bf J}}

\def\baw{\overline w}
\def\baz{\bar z}
\def\baj{\bar\jmath}
\def\bjp{\overline{\jmath+1}}
\def\bjpp{\overline{\jmath+2}}
\def\bet{\bar\eta}
\def\bag{\bar\gam}
\def\baB{\bar B}

\def\zhat{{\bf{\hat z}}}
\def\nhat{{\bf{\hat n}}}

\def\tcrv#1#2{\begin{pmatrix}#1&#2\end{pmatrix}}
\def\tbtm#1#2#3#4{\begin{pmatrix}#1&#2\\#3&#4\end{pmatrix}}
\def\tccv#1#2{\begin{pmatrix}#1\\#2\end{pmatrix}}

\def\fcl{{\rm fcl}}
\def\wey{{\mathcal{W}}}
\def\fop{{\mathcal{D}}}
\def\fopd{\fop_{\rm disc}}
\def\fopc{{\mathcal{C}}}
\def\GSK{\Gam_{\rm SK}}

\def\jtil{\tilde{\jmath}}
\def\Fe8{Fe$_8$}
\def\Mn12{Mn$_{12}$}

\def\am{{\cal L}}
\def\ylm{Y_{\ell m}}
\def\Ylm{{{\cal Y}}_{\ell m}}

\def\wlm{\Phi^{W}_{\ell m}}

\begin{document}

%\twocolumn[
%\hsize\textwidth\columnwidth\hsize\csname@twocolumnfalse\endcsname

\title{The Semiclassical Coherent State Propagator in the Weyl Representation}
\author{Carol Braun$^{\dagger}$}
\author{Feifei Li$^{\dagger\dagger}$}
\author{Anupam Garg}
%\author{carol, feifei, mike, self}
\email[e-mail address: ]{agarg@northwestern.edu}
\affiliation{Department of Physics and Astronomy, Northwestern University,
Evanston, Illinois 60208}
\author{Michael Stone}
\affiliation{Department of Physics, University of Illinois at Urbana-Champaign,
1110 W.~Green St., Urbana, Illinois 61801}

\date{\today}

\begin{abstract}
It is shown that the semiclassical coherent state propagator takes its simplest form when the
quantum mechanical Hamiltonian is replaced by its Weyl symbol in defining the classical action, in that
there is then no need of a Solari-Kochetov correction. It is also shown that such a correction exists if
a symbol other than the Weyl symbol is chosen, and that its form is different depending on the symbol chosen.
The various forms of the propagator based on different symbols are shown to be equivalent provided
the correspondingly correct Solari-Kochetov correction is included. All these results are shown for both
particle and spin coherent state propagators. The global anomaly in the fluctuation determinant is further
elucidated by a study of the connection bewteen the discrete fluctuation determinant and the discrete Jacobi
equation.
\end{abstract}

\maketitle

%\newpage
\section{Introduction}

Coherent-state path integrals for spin and for linear position and momentum degrees of freedom (and related
phase-space path integrals) have been the subject of much study for over three decades
now~\cite{kla78,kla79,kur80,jev79,nie88,kur92,sol87,koc95,vie95},
both for intrinsic reasons, and for their semiclassical limit, where they find application to many practical
problems. Their mathematical subtleties have, however, prevented their widespread use, in contrast to the
Feynman-position space path integral. For example, where in the Feynman integral the paths must be continuous
but need not be differentiable, in the coherent-state case the paths need not even be continuous.

In more recent years, steady progress has been made in understanding the semiclassical limit of such path
integrals~\cite{enz86,bel97}, and in Ref.~\onlinecite{spg00} it was shown that when the so-called Solari-Kochetov
(SK) correction is included, the resulting propagator in the spin case is free of $j$ versus $j+\tshf$ arbitrariness,
has the correct short-time behavior to $O(T^2)$, and is consistent under composition of successive propagators.
In Refs.~\onlinecite{bra07a,bra07b}, this work was extended to coherent-state propagators for many particles and
many spins.

Insights gained from the above work have led to the successful solution for the Bohr-Sommerfeld quantization
rule for spin~\cite{gs04,kur89}, an extension to the instanton calculus~\cite{gkps},
and to a quantitatively correct explanation~\cite{kec02,kec03} of the spin tunneling
spectrum of the magnetic molecule Fe$_8$(tacn)$_6$~\cite{wer99}. Still, the answers for the propagators are obtained only
by a careful examination of the discrete path integral, and casual application of methods developed for the continuous-time 
Feynman path integral is fraught with errors. A continuous-time approach was adopted in Ref.~\onlinecite{spg00}, where
it was found the path integral for the fluctuation determinant suffers from a global anomaly. The resolution of this
problem again requires a careful examination of the discrete path integral, and it is shown that the anomaly is absent in
a special gauge, whereby the Solari-Kochetov correction is automatically included.

While these successes mean that the coherent-state path integral is no longer the heffalump it once was, there is still
some ambiguity in its conception. In particular, while it has long been known that the symbol (or c-number function)
that plays the role of the Hamiltonian in the classical action is not unique~\cite{lie73}, how this nonuniqueness plays
out in the final answer for the semiclassical propagator has not been properly explored. It is the purpose of this paper
to do so, and in the process elucidate the nature of the SK correction further. We will show that the SK correction is
different, depending on the particular Hamiltonian symbol employed, but that the final answer is independent of this
choice. Further, the final answer is best written using the Weyl symbol. As will  become clear, this means that the
formal, continuous-time coherent-state path integral is not only formal, it is also ambiguous. To give it meaning, one
must return to the discrete path integral every time.

Hints that the difficulties of coherent-state path integrals could be related to symbol-choice
ambiguities (or what is the same thing, operator ordering ambiguities) may be seen in
Refs.~\onlinecite{enz86,bel97,kur89,gs04}. Further, Kochetov~\cite{koc98} and Pletyukhov~\cite{ple04} noted that the
SK correction could be written as the difference between the Q symbol and the Weyl symbol for the Hamiltonians, so that
if, in contrast to previous papers which had employed the Q symbol, one employed the Weyl symbol in constructing the
classical action, there would be no SK correction. Pletyukhov showed this result for a system with position and momentum
degrees of freedom in generality, and for spin degrees of freedom within the Holstein-Primakoff approximation.
The absence of the SK correction when one employs the Weyl symbol for
the Hamiltonian suggests at first that it is simply clumsy to have worked with the Q symbol, and that if one uses the
Weyl symbol from the outset, the correction will simply not arise in the first place. If true, this would be a nontrivial
result since, as shown in Ref.~\olc{spg00}, the correction
arises from a global anomaly in the fluctuation determinant, and it is not
clear how a change in the way the extremal action is expressed affects the fluctuations. Indeed, it is not clear how one
would do the calculation with a general symbol in the first place. With this in mind, we calculate the propagator for
particles in the P representation, following closely the derivation based on the Q representation in Sec.~II of
Ref.~\onlinecite{bra07a}. Although the resulting change in the discrete path integral is seemingly minor, it leads to a
nontrivial change in the final answer, and the SK correction now appears with the opposite sign. We then show that both
this answer and the one from the Q representation are equivalent to each other, and to that for Weyl representation.
We also show the analogous result for the spin case. In this we corroborate Pletyukhov, but we do not limit ourselves to
the Holstein-Primakoff approximation, so our proof is completely general. The Weyl symbol for operators based on position and
momentum degrees of freedom is of course classic~\cite{weyl}, but an analogous one exists for spin degrees of freedom
too~\cite{Strato56,Bayen78,Girish81,vgb89,lbg13}, although it is less well known.

%Discuss here why the abscence ot the SK correction is a big deal.

The plan of the paper is as follows. We present the results for the Weyl-representation propagators for both particles and spin
in the next section. This section also serves to introduce our notation, and to define principal terms. The P-representation
calculation is done in \Sno{partP}, and the equivalence of the Q-, P-, and Weyl-symbol-based answers is shown in \Sno{PQW_equiv}.
The propagator for spin in the Weyl representation is derived in \Sno{spinW}. In Secs.~\ref{fcl} and \ref{tridiag}, we turn to
an examination of the continuum and discrete fluctuation determinants and their connection with the corresponding Jacobi
equations with the goal of shedding more light on the global anomaly. Finally, in \Sno{partW} we consider what happens
when we try and evaluate the propagator for particles by working directly with the discrete action using the Weyl representation.
Some essential facts about the Weyl representation (for both particles and spin) are collected in Appendix~\ref{symbols}.

\section{Principal results}
\label{results}

\subsection{Propagator for particles}
\label{part_prop}

For a particle with linear momentum $p$, coordinate $q$, and arbitrary Hamiltonian $\ham$, the propagator is defined as
\beq
K(\baz_f, z_i; T) = \mel{\baz_f}{e^{-i\ham T}}{z_i}.  \label{def_K_part}
\eeq
We have introduced here (unnormalized) harmonic-oscillator-based coherent states,
\beq
\ket{z} = e^{z\adag} \ket{0}, \quad \bra{\baz} = \bra{0} e^{\baz a}, 
   \label{def_cs_part}
\eeq
with $\ket{0}$ and $\bra{0}$ being the normalized ket and bra for the ground state, and $a$ and $\adag$ being the
annihilation and creation operators. In \eno{def_K_part}, $z_i$ and $\baz_f$ are arbitrary complex numbers.

The Weyl form of the semiclassical approximation to the propagator $K$ is
\beq
K^W(\baz_f, z_i; T)
   = \left(i \frac{\ptl^2}{\ptl\baz_f \ptl z_i} S^W(\baz_f, z_i; T) \right)^{1/2}
        \exp \left[iS^W(\baz_f, z_i; T) \right].  \label{ans_KW}
\eeq
Here, the classical action $S^W$ is given by
\beq
iS^W(\baz_f, z_i; T)
    = \hf\bigl[\baz_f z(T) + \baz(0) z_i \bigr]
       + \int_0^T \left[\frac{\dot\baz z - \baz \dot z}{2} - i H^W\bigl[\baz(t), z(t)\bigr] \right]\, dt,
    \label{ans_SW}
\eeq
with $H^W(\baz, z)$ being the Weyl symbol for the Hamiltonian, and $z(t), \baz(t)$ being the solution to the
classical equations of motion
\bea
\frac{d\baz}{dt} &=& i \frac{\ptl H^W}{\ptl z}, \label{eom_W1}\\
\frac{dz}{dt} &=& -i \frac{\ptl H^W}{\ptl \baz},
   \label{eom_W}
\eea
with the boundary conditions $z(0) = z_i$, $\baz(T) = \baz_f$.

We discuss the Weyl symbol $H^W$ at greater length in Appendix \ref{symbols}. For now it suffices to recall
the common textbook definition: $H^W$ is the c-number function obtained by symmetrizing $\ham$ in
$a$ and $\adag$ and then replacing these operators by the c-numbers $z$ and $\baz$ respectively.

The central point of the result (\ref{ans_KW}) is that it has no Solari-Kochetov correction. For comparison,
when we use the Q symbol, the semiclassical approximation to $K$ takes the form~\cite{spg00,bra07a}
\beq
K^Q(\baz_f, z_i; T)
   = \left(i \frac{\ptl^2 S^Q}{\ptl\baz_f \ptl z_i} \right)^{1/2}
        \exp \left[iS^Q(\baz_f, z_i; T)  +\frac{i}{2} \int_0^T A^Q(t)\, dt \right].  \label{ans_KQ}
\eeq
The action $S^Q$ is given by Eqs.~(\ref{ans_SW})--(\ref{eom_W}) with the superscript $W$ replaced by $Q$ everywhere,
with, additionally, $H^Q$, the Q representation of the Hamiltonian~\cite{no_Q_in07}, being defined by
\beq
H^Q(\baz_{j+1}, z_j) = \frac{\mel{\baz_{j+1}}{\ham}{z_j}}{\olap{\baz_{j+1}}{z_j}} \label{def_HQ_1}.
\eeq
Lastly,
\beq
A^Q = \frac{\ptl^2 H^Q}{\ptl\baz \ptl z},  \label{ans_AQ}
\eeq
and it is the term containing $A^Q$ which we call the SK correction in \eno{ans_KQ}.

For completeness, we also give the answer for $K$ when we employ $H^P$, the P symbol for the Hamiltonian.
We show in \Sno{partP} that
\beq
K^P(\baz_f, z_i; T)
   = \left(i \frac{\ptl^2 S^P}{\ptl\baz_f \ptl z_i} \right)^{1/2}
        \exp \left[i S^P(\baz_f, z_i; T) -\frac{i}{2} \int_0^T A^P(t)\, dt \right].  \label{ans_KP}
\eeq
All quantites here are the same as in \etwo{ans_KQ}{ans_AQ} with the superscript Q replaced by P, and
$H^P$ defined via
\beq
\ham = \int {d^2z \over \pi} e^{-\baz z} H^P(\baz, z) \kb{z}{\baz}. 
\eeq
The significant point is that the SK correction enters \eno{ans_KP} with a sign opposite to that in
\eno{ans_KQ}.

We will show in \Sno{PQW_equiv} that $K^P$, $K^Q$, and $K^W$ are all equal up to the leading two terms
in an expansion in $\hbar$.

\subsection{Propagator for spin}
\label{spin_prop}

The coherent-state propagator for a spin of magnitude $j$ is defined in parallel with that for particles:
\beq
K(\baz_f, z_i; T) = \mel{\baz_f}{e^{-i\ham T}}{z_i}.  \label{def_K_spin}
\eeq
The Hamiltonian is an arbitrary polynomial in the usual spin operators $J_x$, $J_y$
and $J_z$. It follows that $J^2 = J_x^2 + J_y^2 + J_z^2$ is a constant of motion, which equals
$j(j+1)$ for spin $j$. Further, $\ket{z_i}$ and $\bra{\baz_f}$ are spin coherent states, defined by
\beq
\ket{z} = e^{zJ_-} \ket{j,j}, \quad \bra{\baz} = \bra{j,j} e^{\baz J_+},
    \label{def_cs_spin}
\eeq
with $\ket{j,j}$ being the eigenstate of $J_z$ with eigenvalue $j$, and $J_{\pm} = J_x \pm i J_y$.
Again, the quantities $z_i$ and $\baz_f$ are arbitrary complex numbers which give the stereographic
coordinates of the maximal spin projection direction in space.

The Weyl-symbol-based semiclassical approximation to $K$ is
\beq
K^W(\baz_f, z_i; T)
  = \left(\frac{i} {2\jtil} \frac{\ptl^2 S^W(\baz_f, z_i; T)}{\ptl\baz_f \ptl z_i} \right)^{1/2}
       \exp(iS^W(\baz_f, z_i; T)).
     \label{ans_KW_spin}
\eeq
Here,
\beq
\jtil = j + \tshf.
\eeq
This is reminiscent of the oft-stated prescription for the ``classical" magnitude of the spin. Second,
\beq
i S^W(\baz_f, z_i; T)
    = \jtil \ln\bigl[\bigl(1 + \baz_f z(T)\bigr) \bigl(1 + \baz(0) z_i \bigr)\bigr]
        + \int_0^T dt\, \left[ \jtil\frac{\dot{\baz} z - \baz \dot{z}}{1 + \baz z}
                                                - i H^W(\baz, z) \right]. \label{SW_spin}
\eeq
Third, the path $\bigl(\baz(t), z(t) \bigr)$ that appears in the action is the solution to the classical
equations of motion,
\beq
\frac{d\baz}{dt} = i \frac{(1+\baz z)^2}{2j} \frac{\ptl H^W}{\ptl z}, \quad
\frac{dz}{dt} = -i \frac{(1+\baz z)^2}{2j} \frac{\ptl H^W}{\ptl\baz},
    \label{eom_W_spin}
\eeq
subject to the boundary conditions $z(0) = z_i$, $\baz(T) = \baz_f$.

As the notation suggests, $H^W$ is the Weyl symbol for the Hamiltonian in the above expressions. In contrast
to the particle case, there is no simple analogue of the symmetrization rule for obtaining $H^W$. Rather, it
is defined by the demands that the map from a spin operator $F$ to its Weyl symbol $\Phi^W_F(\baz, z)$ be
linear, covariant under rotations, yield a real c-number function for Hermitian operators, and, most importantly,
obey the traciality condition
\beq
\frac{1}{2j+1} \Tr (FG) = \frac{1}{\pi} \int \frac{d^2z}{(1+\baz z)^2} \Phi^W_F(\baz, z) \Phi^W_G(\baz, z),
\eeq
for any two spin operators $F$ and $G$ and their corresponding Weyl symbols~\cite{trace_pq}. See Ref.~\olc{lbg13}
and references therein for details. A brief catalog of the results most relevant to this paper is given in
Appendix \ref{rev_spin_pqw}.

Again, the significant point is that the form (\ref{ans_KW_spin}) of the propagator needs no
Solari-Kochetov correction. By contrast, the answer based on the Q representation is~\cite{spg00}
\beq
K^Q(\baz_f, z_i; T)
  = \left[i \frac{\bigl(1 + \baz_f z(T)\bigr) \bigl(1 + \baz(0) z_i \bigr)}{2j}
            \frac{\ptl^2 S^Q}{\ptl\baz_f \ptl z_i} \right]^{1/2}
       \exp\left[iS^Q(\baz_f, z_i; T) + \frac{i}{2} \int_0^T \!\!\!A^Q(t)\, dt \right],
   \label{ans_K_spg}
\eeq
with
\bea
&\!\!\!\!\!\!\dst{iS^Q(\baz_f, z_i; T)
  = j \ln\bigl[\bigl(1 + \baz_f z(T)\bigr) \bigl(1 + \baz(0) z_i \bigr)\bigr]
     +\int_0^T dt\, \left[ j\frac{\dot{\baz} z - \baz \dot{z}}{1 + \baz z} - i H^Q(\baz, z) \right]},&
                     \label{SQ_spin} \\[2pt]
&\dst{A^Q(t)
  = \hf \left( \frac{\ptl}{\ptl\baz} \frac{(1+\baz z)^2}{2j} \frac{\ptl H^Q}{\ptl z}
              + \frac{\ptl}{\ptl z} \frac{(1+\baz z)^2}{2j} \frac{\ptl H^Q}{\ptl\baz} \right),}&
\eea
and
\beq
H^Q(\baz, z) = \frac{\mel{\baz}{\ham}{z}}{\olap{\baz}{z}}.
\eeq
The quantity $A$ is the integrand of the Solari-Kochetov term, and $H^Q$ is the Q symbol for $\ham$. The classical
path obeys \eno{eom_W_spin} with $H^Q$ in lieu of $H^W$.

We do not bother to write $K^P$ explicitly; it would be completely parallel to \eno{ans_K_spg}, with the
sign of the SK term reversed. This follows from what we do in \Sno{partP} and its extension to spin as indicated
in that section.

\section{Coherent-state propagators for particles}
\label{partP}

In this section, we consider the coherent-state propagator for particles. We show in \Sno{setup_pi} that
the path integral for the propagator is not unique, and illustrate this by giving three different
expressions for it. The first two are based on the Q and P symbols for the Hamiltonian, while the third
uses alternating Q and P symbols. The semiclassical propagator studied in
Refs.~\olc{sol87,koc95,vie95,spg00,bra07a,bra07b} is the one
based on the Q-symbol expression, and it contains the original SK correction.
We will calculate the P-symbol-based propagator in \Sno{successive}, where it will be seen that the SK
correction arises with the opposite sign from that when the Q symbol is employed. The calculation
starting from the mixed P-Q expression will be given in \Sno{alternate}.

\subsection{Setting up the path integral}
\label{setup_pi}

We expect on general grounds that, in any semiclassical approximation, $K \sim \exp(iS)$, where $S$ is the action 
for the classical path running from the initial state to the final state. However, this degree of approximation
(analogous to the eikonal approximation in the WKB method) is too crude, for it ignores the conservation of
probability. For that, one must include the next term in an expansion in powers of $\hbar$. This correction
generally takes the form of a pre-exponential factor $\sim(\ptl^2 S/\ptl \baz_f \ptl z_i)^{1/2}$, but of course we
must find it more precisely. It is clear, however, that one must also calculate the exponent or eikonal correct
to the leading {\it two\/} orders in an expansion in $\hbar$. All these points are well known, but we dwell on
them because the next-to-leading-order term is the source of all the trouble in all coherent-state-based
semiclassical propagators and of Solari-Kochetov corrections in particular. Finding this term correctly is important
as it is the one that assures conservation of probability.

To calculate $K$, we divide the interval $T$ into $M$ slices of width $\Dta$ each:
\beq
\Dta = T/M,
\eeq
where $M \gg 1$, so that $\Dta$ is infinitesimal and an expansion in $\Dta$ is permissible. We then write
\beq
e^{-i\ham T} = e^{-i\ham\Dta} e^{-i\ham\Dta} \cdots e^{-i\ham\Dta} \quad (M {\rm\ factors}).
 \label{t_slice}
\eeq
Next, we insert a resolution of unity between every pair of adjacent factors in \eno{t_slice}. Correct to order
$\Dta$, the propagator for one time slice is now evaluated as
\beq
\mel{\baz_{j+1}}{e^{-i\ham\Dta}}{z_j}
  = \olap{\baz_{j+1}}{z_j} \exp\bigl(-i\Dta H^Q(\baz_{j+1}, z_j) + O(\Dta^2)  \bigr),
\eeq
where~\cite{no_Q_in07}
\beq
H^Q(\baz_{j+1}, z_j) = \frac{\mel{\baz_{j+1}}{\ham}{z_j}}{\olap{\baz_{j+1}}{z_j}}. \label{def_HQ}
\eeq
Inserting the explicit expressions for overlaps such as $\olap{\baz_{j+1}}{z_j}$, we obtain
\beq
K(\baz_f, z_i; T) \apx \biggl[\prod_{j=1}^{M-1} \int {d^2z_j \over \pi} \biggr] \exp{(iS^Q_{\rm disc})},
\eeq
where $S^Q_{\rm disc}$, the discrete action, is
\bea
iS^Q_{\rm disc}
  &=&  (\baz_M z_{M-1} - \baz_{M-1} z_{M-1})
   +  (\baz_{M-1} z_{M-2} - \baz_{M-2} z_{M-2}) + \cdots \nnu \\
  &\ & {}+ (\baz_2 z_1 - \baz_1 z_1) + \baz_1 z_0 - i\Dta \sum_{j=0}^{M-1} H^Q(\baz_{j+1}, z_j).
\label{S_Qrep}
\eea
Here, $\baz_M \equiv \baz_f$, and $z_0 \equiv z_i$, and it should be observed that $S^Q_{\rm disc}$ does not depend on
$z_M$ and $\baz_0$ for the simple reason that no such variables have been defined in the first place.

We obtain a different expression for $K$ based on the P symbol, if, again correct to order $\Dta$, we write the $j$th
factor from the right in the string (\ref{t_slice}) as
\beq
e^{-i\ham\Dta} = \int {d^2z_j \over \pi} e^{-\baz_j z_j - i\Dta H^P(\baz_j, z_j)} \kb{z_j}{\baz_j}.
\eeq
Carrying out this substitution, we obtain
\beq
K(\baz_f, z_i; T) \apx \biggl[\prod_{j=1}^M \int {d^2z_j \over \pi} \biggr] \exp{(iS^P_{\rm disc})},
\eeq
where $S^P_{\rm disc}$ is another discrete action, given by
\bea
iS^P_{\rm disc}
  &=&  (\baz_{M+1} z_M - \baz_M z_M)
             +  (\baz_M z_{M-1} - \baz_{M-1} z_{M-1}) + \cdots \nnu\\
  &\ & {}+ (\baz_2 z_1 - \baz_1 z_1) + \baz_1 z_0 - i\Dta \sum_{j=1}^M H^P(\baz_j, z_j).
\label{S_Prep}
\eea
Again, $\baz_{M+1} \equiv \baz_f$, $z_0 \equiv z_i$, and variables $z_{M+1}$ and $\baz_0$ do not exist,
never having been defined.

\Eno{S_Prep} differs from \eno{S_Qrep} in two ways. The first is that we now have $M$ integrations instead of $M-1$.
Since we are eventually going to let $M \to \infty$, this change is insignificant. The second difference is that
$H^P$ is evaluated at $\baz_j$
and $z_j$ in the $j$th slice, whereas $H^Q$ is evaluated at $\baz_{j+1}$ and $z_j$. When we evaluate the extremal value of
the action, we do so on a path where $\baz_{j+1} - \baz_j = O(\Dta)$, so the second change would also appear to be
inconsequential. Yet it is on precisely this difference that everything will pend, for it affects the essential properties
of the two fluctuation operators vis-a-vis their self-adjointness, or lack thereof.

We obtain yet another expression for $K$ if, instead of using all P's or all Q's, we alternate between the two.
Let us consider the first two time steps starting with the state $\ket{z_0}$ ($z_0 \equiv z_i$). We approximate
propagation in the first step via $H^P$, i.e., we write
\beq
e^{-i\ham\Dta} \ket{z_0}
     \apx \int \frac{d^2z_1}{\pi} e^{-\baz_1 z_1} e^{-i\Dta H^P(\baz_1, z_1)}
                \ket{z_1}\olap{\baz_1}{z_0}.
\eeq
We propagate across the next time step by evolving the integrated-over state $\ket{z_1}$ which appears above 
via $H^Q$, i.e., we write
\bea
e^{-i\ham\Dta} \ket{z_1}
     &=& \int \frac{d^2z_2}{\pi} e^{-\baz_2 z_2} \ket{z_2}
          \mel{\baz_2}{e^{-i\ham\Dta}}{z_1}, \nnu\\
     &\apx& \int \frac{d^2z_2}{\pi} e^{-\baz_2 z_2} \ket{z_2}
          \Bigl[ e^{\baz_2 z_1} e^{-i\Dta H^Q(\baz_2, z_1)} \Bigr].
\eea
These two steps generate the following part of $iS_{\rm disc}$:
\beq
\baz_2 z_1 - \baz_1 z_1 + \baz_1 z_0
    - i\Dta\bigl(H^Q(\baz_2, z_1) + H^P(\baz_1, z_1)\bigr).
\eeq
We continue in this way, alternating $H^P$ and $H^Q$. The resulting discrete action is
\bea
iS^A_{\rm disc}
  &=&  (\baz_{M+1} z_M - \baz_M z_M)
             +  (\baz_M z_{M-1} - \baz_{M-1} z_{M-1}) + \cdots \nnu\\
  &\ & + (\baz_2 z_1 - \baz_1 z_1) + \baz_1 z_0 
    - i\Dta\sum_{j=1, 3, 5, \ldots} (H^Q(\baz_{j+1}, z_j) + H^P(\baz_j, z_j)\bigr).
\label{S_Arep}
\eea
The superscript A stands for `alternating'.

It is clear that we can use $H^P$ and $H^Q$ in any order, and thus obtain infinitely many discrete
path-integral expressions for $K$. We could also try and write $\mel{\baz'}{e^{-i\ham\Dta}}{z}$ in terms of
the Weyl symbol $H^W$ using \eno{mel_weyl_ker}, extending the set of expressions even more. This
immediately raises the question of how these different expressions will lead to the same semiclassical
answer for $K$. We will address this question for the P representation in the next subsection, for the
mixed P-Q representation in \Sno{alternate}, and for the direct replaceent via the Weyl symbol in
\Sno{map_time_slice}. First, however, let us see what happens if we take the formal continuous-time limit
($\Dta \to 0$), and write the propagator as the path integral
\beq
K_{\rm fcl} = \int[d^2z] \, e^{iS_{\fcl}[\baz, z]},
\eeq
with
\beq
iS_{\rm fcl}
  = \tshf\bigl(\baz_f z(T) + \baz(0) z_i\bigr)
     + \int_0^T dt\, \Bigl[\frac{{\dot\baz} z - \baz{\dot z}}{2} - i H(\baz, z) \Bigr]. \label{S_fcl}
\eeq
Not only is this equation merely formal, it is also meaningless, because $H$ could stand for $H^P$, $H^Q$,
or something else, depending on which discrete path integral one starts with. This lack of meaning explains
why there is an anomaly in the corresponding path integral. If we try and
work with \eno{S_fcl} as was done in Ref.~\olc{spg00}, we will have to first specify what $H(\baz, z)$
means, and, depending on that, the prescription for regulating the global anomaly will be different. This
prescription will have to be obtained by examining the discrete path integral once again, so it seems that one
is best off by working with the discrete form all the way, and eschewing the formal continuous-time
form altogether~\cite{br}.

\subsection{P-representation propagator by integration by successive time slices}
\label{successive}

In this section, we find the particle-case propagator starting with \eno{S_Prep}. We will do this using the
method of Ref.~\olc{bra07a} since this method can be generalized to arbitrarily many particles and to
arbitrarily many spins~\cite{bra07b}. Since these references show how the extension to more than one particle or one
spin is performed, we will show the calculation for one particle only, and leave the obvious generalization to
many particles and many spins to the reader. We use the same notation,
and focus on the changes that arise, so readers may wish to have a copy of Ref.~\olc{bra07a} handy as they read along.

The first step is to find the ``classical" or extremizing path. The equations for this are essentially the same, and formally
identical when we pass to the $\Dta \to 0$ limit. The next step is to expand the action to second order in fluctuations around
the extremizing path. Denoting the deviations in $z_j$ and $\baz_j$ from the classical path by $\eta_j$ and $\bet_j$, the second
variation of the action is
\beq
\dta^2 S^P_{\rm disc}
  = \frac{1}{2!}
      \biggl[\sum_{j=1}^M \Bigl(\eta_j \frac{\ptl}{\ptl z_j} + \bet_j \frac{\ptl}{\ptl\baz_j} \Bigr)\biggr]^2
          S^P_{\rm disc}.
\eeq
In terms of this quadratic form,
the reduced propagator (the quantity multiplying the exponential of the classical action times $i$) is given by
\beq
K^P_{\rm red}(\baz_f, z_i; T) = \biggl[\prod_{j=1}^M \int {d^2\eta_j \over \pi} \biggr] \exp{(i\dta^2S^P_{\rm disc})}.
\eeq
As in Ref.~\olc{bra07a}, most of the derivatives in $\dta^2 S^P_{\rm disc}$ are zero. The exact expressions for the nonzero
coefficients are slightly different, and they are now given by
\bea
&\dst{ D_{jj}
    = -i \frac{\ptl^2 S^P_{\rm disc}}{\ptl z_j^2}
    = i \Dta \frac{\ptl^2}{\ptl z_j^2} H^P(\baz_j, z_j)}, & \label{Dno1}\\
&\dst{ D_{\baj\baj}
    = -i \frac{\ptl^2 S^P_{\rm disc}}{\ptl \baz_j^2}
    = i \Dta \frac{\ptl^2}{\ptl \baz_j^2} H^P(\baz_j, z_j)}, & \\
&\dst{ D_{\baj j}
    = -i \frac{\ptl^2 S^P_{\rm disc}}{\ptl \baz_j \ptl z_j}
    = 1 + i \Dta \frac{\ptl^2}{\ptl \baz_j \ptl z_j} H^P(\baz_j, z_j)}, & \\
&\dst{D_{j \baj}
    = -i \frac{\ptl^2 S^P_{\rm disc}}{\ptl z_j \ptl \baz_j}
    = 1 + i \Dta \frac{\ptl^2}{\ptl z_j \ptl \baz_j} H^P(\baz_j, z_j)}, & \\
&\dst{D_{\bjp j}
    = -i \frac{\ptl^2 S^P_{\rm disc}}{\ptl \baz_{j+1} \ptl z_j}
    = -1},& \\
&\dst{D_{j \bjp}
    = -i \frac{\ptl^2 S^P_{\rm disc}}{\ptl z_j \ptl \baz_{j+1}}
    = -1}.& \label{Dno6}
\eea
Of these the first two are essentially the same as before (i.e., for the Q representation), but the last
four are different. Clearly, $D_{\baj j} = D_{j \baj}$ and $D_{\bjp j} = D_{j \bjp}$.

The procedure at this point is to carry out the integrals time slice by successive time slice, and step three is to isolate the
quantities that appear in the integral at the $j$th slice. We write this integral as
\beq
\int \frac{d^2\eta_j}{\pi}
   \exp\left[- \hf \tcrv{\bet_j}{\eta_j} G_j \tccv{\eta_j}{\bet_j}
              + {\tilde V}_j \tccv{\eta_j}{\bet_j} + \tcrv{\bet_j}{\eta_j} V_j \right],
\eeq          
just as Eq.~(2.34) in Ref.~\olc{bra07a}. The quantities $\eta_j$ and $\bet_j$ are the deviations in $z_j$ and $\baz_j$
from the classical path, ${\tilde V}_j$ and $V_j$ are row and column vectors given by
\bea
&\dst{{\tilde V}_j
    = -\hf \tcrv{\bet_{j+1}}{\eta_{j+1}} \tbtm{D_{\bjp j}}{0}{0}{0}
    = \hf \bet_{j+1} \tcrv{1}{0}}&, \\
&\dst{V_j
    = -\hf \tbtm{0}{0}{0}{D_{j \bjp}} \tccv{\eta_{j+1}}{\bet_{j+1}}
    = \hf \tccv{0}{1} \bet_{j+1}}&,
\eea
and $G_j$ is a $2\times 2$ matrix that will be determined recursively. To avoid confusion with the time-slice labels, we
label its elements with the letters ``u" and ``d" (for ``up" and ``down"), thus:
\beq
G_j = \tbtm{G_{j,uu}}{G_{j,ud}}{G_{j,du}}{G_{j,dd}}.
\eeq

The fourth step is to shift $\eta_j$ and $\bet_j$ so as to complete the square, and perform the integration
for the $j$th slice. The shifts are given by
\bea
&\dst{\tccv{\gam_j}{\bag_j}
    = G_j^{-1} \tccv{0}{1} \bet_{j+1}}&,\\
%    = \frac{1}{\det G_j} \tccv{G_{j,ud}}{-G_{j,uu}}} \bet_{j+1}&, \\
&\dst{\tcrv{\bag_j}{\gam_j}
    = \bet_{j+1} \tcrv{1}{0} G_j^{-1}}&.
%    = \frac{\bet_{j+1}}{\det G_j} \tcrv{-G_{j,dd}}{G_{j,ud}}}&.
\eea
This leads, as before, to the consistency condition
\beq
G_{j,uu} = G_{j,dd}. \label{Gj_consistent}
\eeq
This condition holds for $j=1$, since
\beq
G_1 = \tbtm{D_{{\bar1}1}}{D_{\bar1\bar1}}{D_{11}}{D_{1\bar1}}
\eeq
and $ D_{{\bar1}1} = D_{1\bar1}$. We shall see from the recursion found below that it
holds for all $j$. The integral gives an overall factor of $(\det G_j)^{-1/2}$, and a residual term in the
exponent from completing the square,
\beq
\hf \tcrv{\bag_j}{\gam_j} G_j \tccv{\gam_j}{\bag_j} = \hf \bet_{j+1}^2 (G_j^{-1})_{ud}. \label{residue_j}
\eeq

Step five is to examine the recursion relation relation for $G_j$ and $\det(G_j)$. \Eno{residue_j} implies that
\beq
G_{j+1} = \tbtm{D_{\bjp\,j+1}}{D_{\bjp\,\bjp}}{D_{j+1\,j+1}}{D_{j+1\,\bjp}}
          - \tbtm{0}{(G_j^{-1})_{ud}}{0}{0}.
\eeq
This shows, first, that the consistency condition (\ref{Gj_consistent}) holds for all $j$. Second,
as in Ref.~\olc{bra07a}, there is no meaningful recursion relation for $\det G_j$, but there is one for the
$ud$ element $G_{ud}$. Since $(G_j^{-1})_{ud} = -G_{j,ud}/\det(G_j)$, this recursion relation is
\beq
G_{j+1,ud} = D_{\bjp\,\bjp} + (\det G_j)^{-1} G_{j,ud},
\eeq
which, a priori, looks different from that in Ref.~\olc{bra07a}. To see its explicit form, we note that
\beq
\det G_j
  = 1 + 2i\Dta \frac{\ptl^2 H^P}{\ptl \baz_j \ptl z_j} 
      - i\Dta \frac{\ptl^2 H^P}{\ptl z_j^2} G_{j,ud} + O(\Dta^2),
    \label{det_Gj}
\eeq 
which along with the expression for $D_{\bjp\,\bjp}$ leads to
\beq
G_{j+1, ud}
     = i\Dta \frac{\ptl^2 H^P}{\ptl \baz_{j+1}^2}
         + \left(1 - 2i\Dta \frac{\ptl^2 H^P}{\ptl \baz_j \ptl z_j} 
                  + i\Dta \frac{\ptl^2 H^P}{\ptl z_j^2} G_{j,ud} +O(\Dta)^2 \right) G_{j,ud}.
     \label{Gud_rr}
\eeq
This explicit form is the same as when we use the Q representation except that instead of $H^P(\baz_j, z_j)$
we have $H^Q(\baz_{j+1}, z_j)$. Writing $j\Dta = t$, and taking the limit $\Dta \to 0$,
\eno{Gud_rr} turns into the Riccati differential equation,
\beq
-i {\dot G}_{ud} = B - 2A G_{ud} + \baB G_{ud}^2,  \label{dot_Gud}
\eeq
which must be solved with the initial condition $G_{ud}(0) = 0$. Here,
\beq
A = \frac{\ptl^2 H^P}{\ptl\baz \ptl z}, \quad
B = \frac{\ptl^2 H^P}{\ptl\baz^2}, \quad
\baB = \frac{\ptl^2 H^P}{\ptl z^2}.
\eeq
The solution to this differential equation is, from Ref.~\olc{bra07a},
\beq
G_{ud}(t) = \frac{1}{\baB(t)}\Bigl( A(t) + i \frac{\dot v}{v} \Bigr). \label{ans_Gud}
\eeq
Here, 
\beq
v(t) = \frac{\dta \baz(t)}{\dta \baz(0)},
\eeq
which is a Jacobi field that describes how the classical trajectory for $\baz(t)$ changes upon a change in
the initial value of $\baz(0)$ while holding $z(0)$ fixed. In particular,
\beq
v(T) = \left(i \frac{\ptl^2}{\ptl\baz_f \ptl z_i} S^P(\baz_f, z_i; T) \right)^{-1}.
\eeq

The quantity of greater interest to us, however, is not $G_{ud}(t)$ but $\det G(t)$ (or $\det G_j$), since
it is this determinant that we pick up from the integration at each time slice. The reduced propagator is
\beq
K^P_{\rm red} = \prod_{j=1}^M (\det G_j)^{-1/2}.
\eeq
Taking logs converts the product into a sum, which turns into an integral in the limit $\Dta \to 0$. We found
$\det G_j$ in \eno{det_Gj}. Hence,
\beq
\ln K^P_{\rm red}
    = -\frac{i}{2} \int_0^T dt\,
         \bigl[ 2A^P(t) - \baB(t) G_{ud}(t) \bigr],
\eeq
which differs from Ref.~\olc{bra07a} in the extra first term, $2A(t)$. Feeding in the solution (\ref{ans_Gud}), we obtain
\beq
\ln K^P_{\rm red} = - \frac{i}{2} \int_0^T A^P(t)\,dt - \hf \ln v(T).
\eeq
Hence, the final answer for the propagator in the semiclassical approximation is, as advertised before,
\beq
K^P(\baz_f, z_i; T)
   = \left(i \frac{\ptl^2 S^P}{\ptl\baz_f \ptl z_i} \right)^{1/2}
        \exp \left[i S^P(\baz_f, z_i; T) -\frac{i}{2} \int_0^T A^P(t)\, dt \right].  \label{ans_KP2}
\eeq

\section{Equivalence of the particle propagator in different representations}
\label{PQW_equiv}

Our goal in this section is to show that \etwo{ans_KP}{ans_KQ} are equivalent, and to write the propagator using
the Weyl representation.

If we look at \etwo{HQ_to_HW}{HP_to_HW} it seems that we can replace $H^Q$ and $H^P$ by $H^W$ in $K^Q$
and $K^P$ and delete the Solari-Kochetov correction. This will turn out to be correct, but there is one
subtlety which we must first mind. The path $\bigl(\baz(t), z(t)\bigr)$ which appears in the action
$S^P$ is obtained by solving the equations of motion (\ref{eom_W1}) and (\ref{eom_W}) but with $H^P$
instead of $H^W$. Let us temporarily denote the path by $\bigl(\baz^P(t), z^P(t)\bigr)$ and the action by
$S^P[\baz^P(t), z^P(t)]$ to emphasize this fact. Let us likewise denote the classical path based
on $H^W$ by $\bigl(\baz^W(t), z^W(t)\bigr)$. Since
\beq
H^W(\baz, z) = H^P(\baz, z)
                 \times \bigl(1 + O(\hbar)\bigr),
\eeq
it follows that
\beq
\baz^W(t) = \baz^P(t)
                 \times \bigl(1 + O(\hbar)\bigr), \quad
z^W(t) = z^P(t)
                 \times \bigl(1 + O(\hbar)\bigr).
\eeq
The action $S^P$ is, however, an extremal value. A small change in the path therefore changes the action
only in second order. That is
\beq
S^P[\baz^W(t), z^W(t)]
   = S^P[\baz^P(t), z^P(t)]
                 \times \bigl(1 + O(\hbar^2)\bigr). \label{insensitive}
\eeq
By the same argument,
\bea
S^W[\baz^W(t), z^W(t)]
   &=& S^P[\baz^W(t), z^W(t)]
                 \times \bigl(1 + O(\hbar^2)\bigr), \\
   &=& S^P[\baz^P(t), z^P(t)]
                 \times \bigl(1 + O(\hbar^2)\bigr).
\eea
Since our goal in calculating the semiclassical propagator is to obtain it correctly up to the first term in
relative order $\hbar$, these changes are beyond the accuracy to which we are working, and may be
neglected. They may be similarly neglected in the prefactor $(\ptl^2 S^P/\ptl\baz_f \ptl z_i)^{1/2}$. To this
order of accuracy, therefore, $K^P = K^W$. By the same argument, $K^Q = K^W$.

\section{Spin propagator in the Weyl representation}
\label{spinW}

In this section we shall give the propagator for spin following the ideas developed in the previous
sections for particles, and notation developed in Ref.~\onlinecite{lbg13}.
Our aim is to obtain \eno{ans_KW_spin}, starting with the previously obtained result, \eno{ans_K_spg}, i.e.,
to rewrite $K^Q$ in terms of the Weyl symbol $H^W(\baz, z)$ in parallel with \Sno{PQW_equiv}. As a
preliminary step, we first discuss the stereographic variables $z$ and $\baz$, which are often more
convenient descriptors of the phase-space sphere than the orientation $\nhat$. If the spherical polar
coordinates of $\nhat$ are taken as $(\tta,\vph)$, then
\beq
z = \tan \tfrac{\tta}{2} e^{i\vph}, \quad
\baz = \tan \tfrac{\tta}{2} e^{-i\vph}.
\eeq
The spin coherent state $\ket{\nhat}$, which is the state with maximum spin projection along
$\nhat$, i.e.,
\beq
\bJ\cdot\nhat \ket{\nhat} = j \ket{\nhat},
\eeq
can clearly be obtained from the state with maximum projection along $\zhat$, i.e., $\ket{j,j}$,
by applying a rotation. When the requisite rotation operator is written in terms of $z$ and $\baz$,
its action on $\ket{j,j}$ can be cast in the form (\ref{def_cs_spin}) up to a multiplicative constant,
i.e.,
\beq
\ket{\nhat} \propto \ket{z}.
\eeq
This result is a proportionality rather than an equality because, as defined in \eno{def_cs_spin}, the
state $\ket{z}$ and its dual bra $\bra{\baz}$ are not normalized; rather
\beq
\olap{\baz}{z'} = (1 + \baz z')^{2j}.
\eeq
The resolution of unity now takes the form
\beq
1 = \frac{2j+1}{\pi} \int \frac{d^2z}{(1+\baz z)^{2j+2}} \kb{z}{\baz}.
\eeq
The benefit of using unnormalized states and $z$, $\baz$ variables is the same as for particle coherent
states: Matrix elements are analytic in $z$ and $\baz$, and we can exploit analyticity to simplify many
calculations.

Next, we note that, by \eno{HW_to_HPQ},
\beq
H^W(\baz, z) = H^Q + \frac{\am^2}{4\jtil} H^Q,
\eeq
where $\am = - i (\nhat \times \grad_{\nhat})$ is the angular momentum operator on phase space (not
the Hilbert space of the states $\ket{j,m}$). In terms of $z$ and $\baz$~\cite{grad2},
\beq
\am^2 = -(1+\baz z)^2 \frac{\ptl^2}{\ptl z \ptl\baz}.
\eeq
Hence, \eno{HW_to_HPQ} may be written as
\beq
H^W(\baz, z)
    = H^Q(\baz, z) - \frac{(1+\baz z)^2}{4\jtil} \frac{\ptl^2 H^Q}{\ptl z \ptl\baz},
\eeq
which is correct up to relative order $1/j$. The next step is to write
\beq
A(t) = A_1(t) + A_2(t),
\eeq
where
\bea
A_1 &=& \frac{(1+\baz z)^2}{2j} \frac{\ptl^2 H^Q}{\ptl z \ptl\baz}, \\
A_2 &=& \frac{(1+ \baz z)}{2j}
            \left(z \frac{\ptl H^Q}{\ptl z} + \baz \frac{\ptl H^Q}{\ptl \baz} \right).
\eea
In $A_1$, we may replace $2j$ by $2\jtil$ in the denominator since $A$ is already
of order $1/j$ relative to $S$, and we do not care about errors of relative order $1/j^2$. Thus,
\beq
A_1(t) = -\frac{\am^2}{2\jtil} H^Q\bigl[\baz(t), z(t) \bigr].
\eeq
For $A_2$, we recast it by using the equations of motion. Thus,
\beq
A_2(t) = -i \frac{\dot{\baz} z - \dot{z} \baz}{1 + \baz z}.
\eeq
The terms $A_1$ and $A_2$ can be combined, respectively, with the second and the first terms in the
integral in \eno{SQ_spin} to yield
\bea
\int_0^T dt\, \left[ j\frac{\dot{\baz} z - \baz \dot{z}}{1 + \baz z} - i H^Q(\baz, z) \right]
   +\frac{i}{2} \int_0^T A(t)\, dt
   &=& \int_0^T dt\, \left[ \jtil\frac{\dot{\baz} z - \baz \dot{z}}{1 + \baz z}
         - i \Bigl(1 + \frac{\am^2}{4\jtil} \Bigr) H^Q(\baz, z) \right] \nnu \\
   &=&\int_0^T dt\, \left[ \jtil\frac{\dot{\baz} z - \baz \dot{z}}{1 + \baz z} - i H^W(\baz, z) \right].
\eea

Next, we observe that the coefficient of the explicit boundary term in the action (\ref{SQ_spin}) can be changed from
$j$ to $\jtil$ by lifting the corresponding term in the prefactor into the exponent. In this way, we obtain
\bea
K^Q(\baz_f, z_i; T)
  &=& \left(\frac{i} {2j} \frac{\ptl^2 S^Q}{\ptl\baz_f \ptl z_i} \right)^{1/2}
        \exp \left(\jtil \ln\bigl[\bigl(1 + \baz_f z(T)\bigr) \bigl(1 + \baz(0) z_i \bigr)\bigr]\right)\nnu\\
  &&\ \ \times \exp\left(\int_0^T dt\, \left[ \jtil\frac{\dot{\baz} z - \baz \dot{z}}{1 + \baz z}
                                                - i H^W(\baz, z) \right] \right).
\eea
We can now make two further changes which only affect our answer to relative order $1/j^2$. First, we can
employ the same argument which led to \eno{insensitive} to replace the path used to calculate the action be the
one based on $H^W$ instead of $H^Q$. Second, we can replace the $j$ in the prefactor by $\jtil$. This gives us
the Weyl-symbol-based propagator, \eno{ans_KW_spin}.

\section{The continuum fluctuation operator and the continuum Jacobi equation}
\label{fcl}

While the action in the formal continuum limit, $S_{\rm fcl}$, is ambiguous it is still useful
to consider the reduced path integral,
\beq
K^{\rm red}_{\fcl}
  = \int [dz]\,[d\baz] 
      \exp \bigl(i\dta^2 S_{\fcl}[\bet, \eta] \bigr),
\eeq
where
\beq
i\dta^2 S_{\fcl}[\bet, \eta]
         = -\frac{i}{2} \int_0^T \tcrv{\bet(t)}{\eta(t)} \fop_{\fcl} \tccv{\eta(t)}{\bet(t)}\,dt,
\eeq
with $\fop_{\fcl}$ being the fluctuation operator
\beq
\fop_{\fcl} = \tbtm{-i\ptl_t + A(t)}{B(t)}{\baB(t)}{i\ptl_t + A(t)},
\eeq
acting on paths that obey the constraints $\eta(0) = \bet(T) = 0$. The ambiguity in the formal continuum
limit shows up as follows. As found in Ref.~\olc{spg00}, because of the global anomaly in the path integral
$K^{\rm red}_{\fcl}$, the operator $\fop_{\fcl}$ has no eignefunctions, not even
one. Thus, $\det \fop_{\fcl}$ cannot be defined as the product of the eigenvalues of $\fop_{\fcl}$. For the
same reason, the standard method for finding this determinant based on solving the associated Jacobi
equation also fails.

To explain the nature of this failure, we now describe the Jacobi-equation-based method. The classical equations
based on the continuum action are
\beq
\frac{dz}{dt} = -i\frac{\ptl H}{\ptl \baz}, \quad
\frac{d\baz}{dt} = i\frac{\ptl H}{\ptl z}.
\eeq
These equations have to be solved with the boundary conditions $z(0) = z_i$, $\baz(T) = \baz_f$. The other boundary
values, $\baz(0)$ and $z(T)$ are not fixed, but emerge from the solution and may thus be regarded as functions of $z_i$,
$\baz_f$, and $T$. If now we use the value of $\baz(0)$ so found and $z_i$ to solve the classical equations of motion
as an initial value problem, we will recover the classical solution for $z(t)$ and $\baz(t)$. If we change the initial
values to $z_i$ and $\baz(0) + \eps$, where $\eps$ is infinitesimal, the solution to the initial value problem will deviate
from the previous one by terms of order $\eps$ in leading order. Denoting the deviations in $z(t)$ and $\baz(t)$ by
$\eps u(t)$ and $\eps v(t)$ respectively, we find the Jacobi equations,
\beq
\tbtm{-i\ptl_t + A(t)}{B(t)}{\baB(t)}{i\ptl_t + A(t)} \tccv{u}{v} = 0, \label{jacobi_eom}
\eeq
with initial conditions $u(0) = 0$, $v(0) = 1$.
(Since we will not need it, we do not bother writing the full Jacobi system allowing for variations in $z_i$ also.)
The Jacobi-based method says that
\beq
\det\fop_{\fcl} = v(T). \label{jac_naive}
\eeq
As found in Ref.~\olc{spg00}, \eno{jac_naive} is incorrect and should be multiplied by an undetermined phase
factor, $e^{i\gam}$. This phase factor is the SK correction, which we now know differs for the P and Q
representations, while $\fop_{\fcl}$ is superficially the same in the two cases.

For completeness, and to enable the reader to understand the answers (\ref{ans_KW}), (\ref{ans_KQ}), and (\ref{ans_KP})
for the propagator, we mention that
\beq
v(T) = \Bigl(\frac{\ptl^2 S}{\ptl\baz_f \ptl z_i}\Bigr)^{-1}.
\eeq
The proof is standard. See, e.g., Sec.~4 of Ref.\,\olc{spg00} or Ref.\,\olc{ag_boulder}.

\section{The discrete fluctuation operator and the discrete Jacobi equation}
\label{tridiag}

In this section, we return to the
discrete path integral and, for both the P and Q representations, examine the fluctuation operator by writing it as a
tridiagonal matrix. The same operator determines the Jacobi equation. We will show that unlike the continuum case,
the determinant of the discrete operator is not simply equal to the solution to the discrete Jacobi
equation but also contains an SK correction. We will see why the correction differs between the two cases (P and Q).
We will further see that the solution to the discrete Jacobi equation tends to the continuum solution in the limit
$\Dta \to 0$. This shows the precise way in which the equality of the Jacobi field and the fluctuation
determinant breaks down in this limit.

This method cannot be extended (at least we do not know how) to more than one particle or spin, but the insights it
provides as described above still make it worth presenting.

\subsection{The source of the SK correction}
\label{fopd}

For either \eno{S_Prep} or \eno{S_Qrep}, we can write (taking the number of intermediate integrations as $M$ in both
cases)
\beq
i\dta^2 S_{\rm disc}[\bet, \eta]
   = - \hf
           \begin{bmatrix}
                \bet_1 & \eta_1 & \bet_2 & \eta_2 & \cdots & \bet_M & \eta_M
           \end{bmatrix}
                \fopd 
           \begin{bmatrix}
                \bet_1 \\ \eta_1 \\ \bet_2 \\ \eta_2 \\ \vdots \\ \bet_M \\ \eta_M
           \end{bmatrix},
\eeq
where $\fopd$ is the discrete fluctuation operator (or matrix)
\beq
\fopd
 = \left[\begin{array}{cc|cc|cc|cc}
        D_{\bar1\,\bar1} & D_{\bar1\,1} & & & & & &\\
        D_{1\,\bar1}     & D_{1\,1}     & D_{1\,\bar2} & & & & &\\
        \hline
                         & D_{\bar2\,1} & D_{\bar2\,\bar2} & D_{\bar2\,2} & & & & \\
                         &              & D_{\bar2\,\bar2} & D_{\bar2\,2} & D_{2\,\bar3} & & & \\
        \hline
       & & & D_{\bar3\,2} & \ddots &      & & \\
       & & & & & \ddots & D_{M-1\,\bar M} & \\
        \hline
       & & & & & D_{\bar M\, M-1} & D_{\bar M\,\bar M} & D_{\bar M\,M} \\
       & & & & & & D_{M\,\bar M}      & D_{M\,M} 
    \end{array}\right].
\eeq
Note that we have reordered the $\eta$'s and $\bet$'s in the row vector, as this makes $\fopd$ a manifestly
symmetric and tridiagonal matrix, albeit complex~\cite{not_herm}.
The reordering leads to $M$ additional factors of $-1$ when the determinant is evaluated, so that
\beq
K_{\rm red} = \bigl[(-1)^M \det\fopd \bigr]^{-1/2}.
\eeq

Next, let us examine the discrete Jacobi equation.
The equations for the classical path that follow from the discrete action can be written as
\beq
\frac{\ptl}{\ptl \baz_j} (-iS_{\rm disc}) = 0, \qquad
\frac{\ptl}{\ptl z_j} (-iS_{\rm disc}) = 0, \quad
    (j = 1, 2, \ldots, M).
\eeq
To derive Jacobi equations from these, we would like to treat $z_0 = z_i$ and $\baz_0$ as initial values.
This, however, is meaningless as there is no such variable as $\baz_0$. Instead, we must take $z_0$ and $\baz_1$
as the initial values, the latter being regarded as determined by $z_i$ and $\baz_f = \baz_{M+1}$. We now keep
$z_i$ unchanged, and let $\baz_1 \to \baz_1 + \eps$. Let us denote the changes induced in $z_j$ and $\baz_j$ by
\beq
\dta z_j = \eps u_j + O(\eps^2), \quad \dta\baz_j = \eps v_j + O(\eps^2).
\eeq
Performing the necessary variations, we obtain,
\beq
 \sum_k \Bigl(u_k \frac{\ptl}{\ptl z_k} + v_k \frac{\ptl}{\ptl\baz_k} \Bigr)
  \tccv{-i\ptl S_{\rm disc}/\ptl \baz_j}{-i\ptl S_{\rm disc}/\ptl z_j}
     = \tccv{0}{0}.
\eeq
Most of the derivatives on the left vanish. When only the nonzero ones are kept, we obtain
\bea
D_{j-1 \baj} u_{j-1} + D_{j \baj} u_j + D_{\baj\baj} v_j &=& 0, \label{D_recur1}\\
D_{jj} u_j + D_{\baj j} v_j + D_{\bjp j} v_{j+1} &=& 0, \label{D_recur2}
\eea
using the definitions of the various $D$'s. [These definitions can be read off by considering only the first member of each
of the equations (\ref{Dno1})--(\ref{Dno6}) and deleting the `P' superscript.]
These equations hold for $j = 1, 2, \ldots, M$, and we must take $u_0 = 0$, $v_1 = 1$. They then determine
$u_j$ for $1 \le j \le M$, and $v_j$ for $2 \le j \le M+1$, all of which are meaningful quantities. We now observe that
we can rewrite them in the form
\beq
\left[\begin{array}{cc|cc|cc|cc}
        D_{\bar1\,\bar1} & D_{\bar1\,1} & & & & & &\\
        D_{1\,\bar1}     & D_{1\,1}     & D_{1\,\bar2} & & & & &\\
        \hline
                         & D_{\bar2\,1} & D_{\bar2\,\bar2} & D_{\bar2\,2} & & & & \\
                         &              & D_{\bar2\,\bar2} & D_{\bar2\,2} & D_{2\,\bar3} & & & \\
        \hline
       & & & D_{\bar3\,2} & \ddots &      & & \\
       & & & & & \ddots & D_{M-1\,\bar M} & \\
        \hline
       & & & & & D_{\bar M\, M-1} & D_{\bar M\,\bar M} & D_{\bar M\,M} \\
       & & & & & & D_{M\,\bar M}      & D_{M\,M} 
    \end{array}\right]
\left[\begin{array}{c}
%\begin{bmatrix}
      v_1 \\ u_1\\ \hline   v_2 \\ u_2 \\ \hline \vdots \\ \vdots \\ \hline  v_M \\ u_M
%      \bet_1 \\ \eta_1 \\ \bet_2 \\ \eta_2 \\ \vdots \\ \bet_M \\ \eta_M
    \end{array}\right]
=
\left[\begin{array}{c}
%\begin{bmatrix}
      0 \\ 0\\ \hline  0 \\ 0 \\ \hline  \vdots \\ \vdots \\ \hline  0 \\ v_{M+1}
%      \bet_1 \\ \eta_1 \\ \bet_2 \\ \eta_2 \\ \vdots \\ \bet_M \\ \eta_M
    \end{array}\right]
  \label{full_jacob}
\eeq
We have used the initial condition on $u_0$ ($u_0 = 0$), but not on $v_1$, leaving it as an arbitrary quantity instead. In fact,
the way this equation is written suggests that all $v_j$, $u_j$ ($j = 1, \ldots, M$) are determined in terms of $v_{M+1}$.
By demanding that $v_{M+1}$ must be chosen such that $v_1 = 1$, however, we once again obtain $v_{M+1}$ explicitly.

The matrix that appears in \eno{full_jacob} is of course none other than $\fopd$. By Cramer's rule, therefore,
\beq
v_1 = \frac{\det\fopc}{\det\fopd},
\eeq
where,
\beq
\fopc = 
  \left[\begin{array}{cc|cc|cc|cc}
        0 & D_{\bar1\,1} & & & & & &\\
        0 & D_{1\,1}     & D_{1\,\bar2} & & & & &\\
        \hline
        0                & D_{\bar2\,1} & D_{\bar2\,\bar2} & D_{\bar2\,2} & & & & \\
        0                &              & D_{\bar2\,\bar2} & D_{\bar2\,2} & D_{2\,\bar3} & & & \\
        \hline
       \vdots & & & D_{\bar3\,2} & \ddots &      & & \\
       \vdots & & & & & \ddots & D_{M-1\,\bar M} & \\
        \hline
        0 & & & & & D_{\bar M\, M-1} & D_{\bar M\,\bar M} & D_{\bar M\,M} \\
        v_{M+1} & & & & & & D_{M\,\bar M}      & D_{M\,M} 
    \end{array}\right].
\eeq
To evaluate $\det\fopc$, we expand it by the first column, obtaining,
\beq
\det\fopc = - v_{M+1} \det
  \left[\begin{array}{c|cc|cc|cc}
        D_{\bar1\,1} & & & & & &\\
        D_{1\,1}     & D_{1\,\bar2} & & & & &\\
        \hline
        D_{\bar2\,1} & D_{\bar2\,\bar2} & D_{\bar2\,2} & & & & \\
        &  D_{\bar2\,\bar2} & D_{\bar2\,2} & D_{2\,\bar3} & & & \\
        \hline
        & & D_{\bar3\,2} & \ddots &      & & \\
        & & & & \ddots & D_{M-1\,\bar M} & \\
        \hline
         & & & & D_{\bar M\, M-1} & D_{\bar M\,\bar M} & D_{\bar M\,M}
    \end{array}\right].
\eeq
The matrix that remains is lower triangular, so its determinant is just the product of the diagonal entries. Anticipating
future minus signs, we define the quantity
\beq
\Gam_{\rm SK} = (-1)^{M-1} \prod_{j=1}^M D_{\baj j} \prod_{j=2}^M D_{j-1 \baj},
\eeq
in terms of which
\beq
\det\fopc = (-1)^M v_{M+1} \Gam_{\rm SK}
\eeq
and
\beq
v_1 = (-1)^M \frac{\Gam_{\rm SK}}{\det\fopd} v_{M+1}.
\eeq
Setting $v_1 =1$, and rearranging, we get
\beq
(-1)^M \det\fopd = \GSK v_{M+1}. \label{det_jac_disc}
\eeq
This is the correct discrete replacement of \eno{jac_naive}. The factor $\GSK$ need not be unity, in which case
we have a nonzero SK correction.

The final step is to take the $\Dta \to 0$ limit of \eno{det_jac_disc}. By definition, the left hand side turns into
the continuum fluctuation determinant, and its inverse square root will give us $K_{\rm red}$. It remains to see what
happens to $v_{M+1}$ and $\Gam_{\rm SK}$ on the right hand side. It is simplest to do this separately for the
P and Q representations. Before turning to this, however, it pays to rewrite the general Jacobi equations,
(\ref{D_recur1}) and (\ref{D_recur2}), as the $2\times 2$ matrix recursion relation,
\beq
\tbtm{-D_{j\baj}}{0}{D_{jj}}{D_{\bjp j}}
   \tccv{u_j}{v_{j+1}}
  = \tbtm{D_{j-1 \baj}}{D_{\baj\baj}}{0}{-D_{\baj j}}
      \tccv{u_{j-1}}{v_j}, \quad (j=1, 2, \ldots, M),
    \label{dj}
\eeq
with the initial conditions $u_0 =0$, $v_1 = 1$.

It is of course also possible to evaluate $\det\fopd$ by writing a recursion relation for successive diagonal
subdeterminants of $\fopd$ (as may be done for any tridiagonal matrix). The SK corrections can then be obtained by
examining the relationship of this recursion relation to \eno{dj}. We shall not follow this route.

\subsection{Application to P representation}
\label{discP}

Let us consider the discrete Jacobi equation first. We feed the explicit values of the $D$'s from
Eqs.(\ref{Dno1})--(\ref{Dno6}) into \eno{dj}, and abbreviate
\beq
  A(\baz_j, z_j) = A_j, \quad
  B(\baz_j, z_j) = B_j, \quad
  \baB(\baz_j, z_j) = \baB_j.
\eeq
We find that
\beq
\tbtm{-(1+i\Dta A_j)}{0}{i\Dta\baB_j}{-1}
   \tccv{u_j}{v_{j+1}}
  = \tbtm{-1}{i\Dta B_j}{0}{1+i\Dta A_j}
      \tccv{u_{j-1}}{v_j}.
\eeq
Solving for the column vector on the left, and dropping terms of $O(\Dta^2)$, we obtain
\beq
\tccv{u_j}{v_{j+1}}
  = \tbtm{1 - i\Dta A_j}{-i\Dta B_j}{i\Dta\baB_j}{1+i\Dta A_j}
      \tccv{u_{j-1}}{v_j}.
   \label{uv_disc_P}
\eeq
It is immediately apparent that in the continuum limit ($\Dta \to 0$, $M \to \infty$, with $\Dta M =T$ fixed),
this recursion will turn into the continuous-time Jacobi equation (\ref{jacobi_eom}). Since
the initial conditions are also identical, it follows that
\beq
  \lim_{\Dta \to 0} v_{M+1} = v(T).
\eeq

The second step is to evaluate $\GSK$. We have,
\beq
D_{\baj j} = 1 + i\Dta A_j, \quad D_{j-1 \baj} = -1.
\eeq
Hence,
\beq
\GSK = (-1)^{M-1} \times \prod_{j=1}^{M} (1+i\Dta A_j) \times (-1)^{M-1}
     \apx \exp\bigl(i\Dta \sum_j A_j \bigr).
\eeq

Collecting together the above results, we find that
\beq
\lim_{\Dta \to 0} (-1)^M \det\fopd = \exp \Bigl(i\int_0^T A(t)\,dt\Bigr) v(T),
\eeq
so that
\beq
K^P_{\rm red} = 
    \exp \Bigl(-\frac{i}{2}\int_0^T A(t)\,dt\Bigr) \bigl[v(T)\bigr]^{-1/2}.
\eeq
The extra exponential factor is the SK correction, and we see that we have the correct sign for it.

\subsection{Application to Q representation}
\label{discQ}

We now repeat the previous subsection's arguments for the Q representation. The relevant $D$ coefficients are
given in Eqs.(2.16)--(2.21) of Ref.~\olc{bra07a}, and we redisplay them here for ready reference:
\bea
&D_{jj} = i\Dta \baB_j, \quad D_{\baj\baj} = i\Dta B_j,& \nnu\\
&D_{\baj j} = D_{j\baj} = 1,& \\
&D_{\bjp j} = D_{j \bjp} = -1 + i\Dta A_j,& \nnu
\eea
where now,
\beq
A_j = A(\baz_j, z_{j-1}), \quad
B_j = B(\baz_j, z_{j-1}), \quad
\baB_j = \baB(\baz_{j+1}, z_j).
\eeq
The discrete Jacobi equation now reads
\beq
\tbtm{-1}{0}{i\Dta\baB_j}{-1+i\Dta A_{j+1}}
   \tccv{u_j}{v_{j+1}}
  = \tbtm{-1+i\Dta A_j}{i\Dta B_j}{0}{-1}
      \tccv{u_{j-1}}{v_j}.
\eeq
Again we solve for $u_j$ and $v_{j+1}$ to $O(\Dta)$. In the process, we also replace the $\Dta A_{j+1}$ term by
$\Dta A_j$, since the difference is $O(\Dta^2)$. In this way, we get
\beq
\tccv{u_j}{v_{j+1}}
  = \tbtm{1 - i\Dta A_j}{-i\Dta B_j}{i\Dta\baB_j}{1+i\Dta A_j}
      \tccv{u_{j-1}}{v_j},
   \label{uv_disc_Q}
\eeq
which is formally the same as in the P case. For the same reasons as given there, we again get
\beq
  \lim_{\Dta \to 0} v_{M+1} = v(T).
\eeq

Next, for $\GSK$, we have
\beq
\GSK = (-1)^{M-1} \times (1)^M \times \prod_{j=1}^{M-1} (-1+i\Dta A_j)
     \apx \exp\bigl(-i\Dta \sum_j A_j \bigr).
\eeq
It follows that,
\beq
\lim_{\Dta \to 0} (-1)^M \det\fopd = \exp \Bigl(-i\int_0^T A(t)\,dt\Bigr) v(T),
\eeq
and
\beq
K^Q_{\rm red} = 
    \exp \Bigl(\frac{i}{2}\int_0^T A(t)\,dt\Bigr) \bigl[v(T)\bigr]^{-1/2}.
\eeq
The extra exponential factor is the SK correction for $K^Q$. We draw the reader's attention to the sign.

\section{Direct evaluation of the particle propagator in the Weyl representation}
\label{partW}

Our goal in this section is to try and evaluate the propagator using the Weyl representation for $\ham$ from the very start.
One way to try and do this is to write the infinitesimal time-evolution operator $e^{-i\ham\Dta}$ in terms of $H^W$ using the
Weyl kernel $\wey(\baz,z)$ as in \eno{op_to_weyl}. We shall see that this way does not work. The other way, which does work,
is to alternate P and Q representations, building on \eno{S_Arep}.

\subsection{Mapping via Weyl kernel}
\label{map_time_slice}

First, let us use \eno{op_to_weyl} to write the infinitesimal time-evolution operator as
\beq
e^{-i\ham\Dta} = \int \frac{d^2z}{\pi} \bigl[e^{-i\ham\Dta} \bigr]_{\rm WS}\, \wey(\baz, z),
\eeq
where by $[X]_{\rm WS}$ we mean the Weyl symbol of the operator $X$. Using \eno{mel_weyl_ker}, we may write the
propagator for one time slice as
\beq
\mel{\baz_2}{e^{-i\ham\Dta}}{z_1}
  = 2e^{\baz_2 z_1} \int \frac{d^2z}{\pi} \bigl[e^{-i\ham\Dta} \bigr]_{\rm WS}\, e^{-2(\baz_2 - \baz)(z_1 - z)}.
\eeq 
To find $\bigl[e^{-i\ham\Dta} \bigr]_{\rm WS}$, we expand $e^{-i\ham\Dta}$ in powers of $\Dta$. The symbols for 1 and $\ham$
are $1$ and $H^W(\baz, z)$, and that for $\ham^2$ is~\cite{moy}
\beq
\bigl[\ham^2\bigr]_{\rm WS} 
  = \bigl[H^W(\baz, z)\bigr]^2
      + \frac{\hbar^2}{4}
         \left[ \frac{\ptl^2 H^W}{\ptl z^2} \frac{\ptl^2 H^W}{\ptl \baz^2}
              - \Bigl(\frac{\ptl^2 H^W}{\ptl \baz \ptl z} \Bigr)^2 \right] + \cdots.
\eeq
The important point here is that the correction is of relative order $\hbar^2$ and not $\hbar$. (We will show the powers  of $\hbar$
relative to the leading term explicitly in this section.)
% We could repeat the calculations of this section almost verbatim
%for the case of spin; aside from extra measure factors of $(1+\baz z)$, the role of $\hbar$ would be played by $1/j$.)
Hence, when we reexponentiate the series, we find
\beq
\bigl[e^{-i\ham\Dta} \bigr]_{\rm WS} = \exp\bigl(-i\Dta H^W(\baz, z) + O(\hbar^2\Dta^2)\bigr).
\eeq
For the time-slice propagator, we get
\beq
\mel{\baz_2}{e^{-i\ham\Dta}}{z_1}
  = 2e^{\baz_2 z_1} \int \frac{d^2z}{\pi} e^{\Phi(\baz, z; \baz_2, z_1)},
\eeq 
with
\beq
\Phi(\baz, z; \baz_2, z_1) = - i\Dta H^W(\baz, z) -2(\baz - \baz_2)(z - z_1) + O(\hbar^2\Dta^2),
\eeq

The natural procedure at this point is to evaluate the integral over $z$ and $\baz$ semiclassically, i.e., by
steepest descents. Let us denote the critical (saddle) point by $\baz_c$ and $z_c$, and partial derivatives by subscripts.
Setting $\Phi_z = \Phi_{\baz} = 0$, we find
\bea
z_c &=& z_1 - \frac{i\hbar^{1/2}}{2} \Dta H^W_{\baz}(\baz_c, z_c) + O(\hbar^{5/2} \Dta^2), \\
\baz_c &=& \baz_2 - \frac{i\hbar^{1/2}}{2} \Dta H^W_{z}(\baz_c, z_c) + O(\hbar^{5/2} \Dta^2),
\eea
where we continue to show powers of $\hbar$ explicitly. Hence, denoting the critical value of $\Phi$ by $\Phi_c$,
we have
\beq
\Phi_c = - i \Dta H^W(\baz_c, z_c) + \tshf \Dta^2\hbar H^W_{z}(\baz_c, z_c) H^W_{\baz}(\baz_c, z_c) + O(\hbar^2\Dta^2),
\eeq
the additional error terms introduced at this step being of order $\hbar^3\Dta^3$.
Since $\baz_c$ and $z_1$ differ from $\baz_2$ and $z_1$ by terms of order $\Dta\hbar^{1/2}$, it is reasonable to perform
a second expansion in $\Dta$. When this is done, we find (the sign of the $\Dta^2$ term should be noted)
\beq
\Phi_c = - i \Dta H^W(\baz_2, z_1)
          - \tshf \Dta^2\hbar H^W_{z}(\baz_2, z_1) H^W_{\baz}(\baz_2, z_1)
            + O(\Dta^2 \hbar^2).
\eeq

The next step is to perform the Gaussian integral over the small deviations from the critical point. Defining
$\eta = z - z_c$, $\bet = \baz - \baz_c$, we have
\beq
\Phi = \Phi_c
   + \hf \Bigl( \Phi_{zz} \eta^2 + 2\Phi_{z\baz}\eta\bet + \Phi_{\baz\baz}\bet^2 \Bigr) + \cdots,
\eeq
with
\bea
\Phi_{zz} &=& -i\Dta \hbar H^W_{zz}, \nnu\\
\Phi_{z\baz} &=& -2 -i\Dta\hbar H^W_{z\baz}, \\
\Phi_{\baz\baz} &=& -i\Dta \hbar H^W_{\baz\baz}. \nnu
\eea
At this point, it is better to leave the derivatives of $H^W$ evaluated at $\baz_c$, $z_c$. The integral gives us the
inverse square root of the determinant of this quadratic form, which equals
\beq
1 - i\frac{\Dta\hbar}{2} H^W_{z\baz} + O(\Dta^2\hbar^2).
\eeq
Evaluating $H^W_{z\baz}$ at $\baz_2$, $z_1$ incurs a further error of the same order, i.e., $\Dta^2\hbar^2$.

Putting all these pieces together, we find, eventually,
\beq
\mel{\baz_2}{e^{-i\ham\Dta}}{z_1}
 = \exp\left( \baz_2 z_1 -i\Dta\bigl[ H^W
                           + \tshf H^W_{z\baz} \bigr]_{\baz_2, z_1} + O(\Dta^2\hbar^2) \right).
\eeq
The order $\Dta$ terms combine to form $H^Q(\baz_2, z_1)$, so that we get, as before,
\beq
\mel{\baz_2}{e^{-i\ham\Dta}}{z_1} \apx \olap{\baz_2}{z_1} e^{-i\Dta H^Q(\baz_2, z_1)},
\eeq
but now we know the order of the terms omitted.

We thus see that it does no good to start with $H^W$, since that entails auxiliary integrations for each time slice, which when
performed will lead to the same fluctuation determinant over the non-auxilliary variables as before. Only the Q and P
representations allow us to dispense with auxilliary integration variables, but then the fluctuation determinant
leads to SK corrections.

\subsection{Alternating P and Q representations}
\label{alternate}

We now start with the discrete action (\ref{S_Arep}) obtained by alternating P and Q representations. This is an obvious
thing to try, since when we determine the extreme or the classical value of the action, the alternating $H^P$'s and
$H^Q$'s will combine to produce $H^W$ as $\Dta \to 0$. The expectation is that the reduced propagator will then be free
of any SK correction.

The only nontrivial part of the calculation is the integration over the fluctuations, which we do by
successive time slices as in \Sno{partP} and Ref.\,\olc{bra07a}, picking up factors of $(\det G_j)^{-1/2}$
at each step. Suppose the step from $j$ to $j+1$ is of type P, and the next step is of type Q.
Then, adopting a notation for the derivatives of $H^P$ and $H^Q$ analogous to that in $D_{jj}$, etc., the
part of $\dta^2 (iS)$ which involves the fluctuations at these steps is
\bea
i\dta^2 S
  &=& \cdots - \frac{i}{2}\Dta H^P_{\baj\baj}\bet_j^2 - (1 + i\Dta H^P_{jj})\eta_j\bet_j \nnu\\
  &&\ -{}\frac{i}{2}\Dta(H^P_{jj} + H^Q_{jj}) \eta_j^2 + (1 -i\Dta H^Q_{j\bjp})\eta_j \bet_{j+1} \nnu\\
  &&\ -{} \frac{i}{2}\Dta H^Q_{\bjp,\bjp} \bet_{j+1}^2 - \eta_{j+1}\bet_{j+1} + \eta_{j+1}\bet_{j+2} + \cdots.
\eea
We do not show the values of $\baz$ and $z$ at which the derivatives of $H^P$ and $H^Q$ are evaluated explicitly;
they are $(\baz_j, z_j)$ and $(\baz_{j+1}, z_j)$, respectively. Since we will eventually let $\Dta \to 0$, these
arguments will take values on the classical path $\baz(t)$, $z(t)$ with $t = j\Dta$.

For the integration at step $j$ (over $\eta_j$ and $\bet_j$), we need the matrix $G_j$ as well as the vectors $V_j$
and ${\tilde V}_j$. We have
\bea
&G_j = \tbtm{1+ i \Dta H^P_{\baj,j}}{G_{j,ud}}{i\Dta(H^P_{jj} + H^Q_{jj})}{1 + i \Dta H^P_{j,\baj}},& \\
&V_j = \hf (1 - i\Dta H^Q_{j,\bjp})\  \bet_{j+1}\tccv{0}{1}, \quad
   {\tilde V}_j = \hf (1 - i\Dta H^Q_{j,\bjp})\  \bet_{j+1}\tcrv{1}{0}.&
\eea
The matrix element $G_{j,ud}$ is of course unknown, having been modified as a result of the previous integration steps.
The integration at this step produces a residual term from completing the square equal to
\beq
\hf \bigl(1 - i\Dta H^Q_{j,\bjp}\bigr)^2 \bet^2_{j+1} \bigl(G_j^{-1}\bigr)_{ud}
= -\frac{1}{2\det G_j} \bigl(1 - 2i\Dta H^Q_{j,\bjp}\bigr) \bet^2_{j+1} G_{j,ud} + O(\Dta^2),
\eeq
and a determinantal factor $(\det G_j)^{-1/2}$. It is apparent that
\beq
\det G_j = 1 + 2i\Dta H^P_{\baj, j} -i\Dta \bigl(H^P_{jj} + H^Q_{jj}\bigr) G_{j,ud}.
\eeq

The result of step $j$ is that the matrix $G_{j+1}$ equals
\beq
G_{j+1} = \tbtm{1}{G_{j+1,ud}}{0}{1},
\eeq
where
\beq
G_{j+1,ud} = i\Dta H^Q_{\bjp,\bjp} + \frac{1}{\det G_j} \bigl(1 - 2i\Dta H^Q_{j,\bjp}\bigr) G_{j,ud}.
\eeq
Further,
\beq
V_j = \hf \bet_{j+2}\tccv{0}{1}, \quad {\tilde V}_j = \hf \bet_{j+2}\tcrv{1}{0}.
\eeq
The integrations at step $j+1$ thus produce a determinantal factor $(\det G_{j+1})^{-1/2} = 1$, and a residual
term from completing the square equal to
\beq
\hf \bet_{j+2}^2 \bigl(G^{-1}_{j+1}\bigr)_{ud} =  -\hf \bet_{j+2}^2 G_{j+1,ud}.
\eeq
Hence,
\bea
G_{j+2,ud}
 &=& i\Dta H^P_{\bjpp,\bjpp} + G_{j+1,ud} \\
 &=& i\Dta H^P_{\bjpp,\bjpp} + i\Dta H^Q_{\bjp,\bjp} \nnu\\
 &&\        + \Bigl(1 - 2i\Dta H^Q_{j,\bjp}
         - 2i\Dta H^P_{\baj, j} +i \Dta \bigl(H^P_{jj} + H^Q_{jj}\bigr) G_{j,ud}\Bigr) G_{j,ud}.
\eea
This is the recursion relation desired, since we have now integrated over the complete repeat pattern. If we
let $\Dta \to 0$, it will be seen that the symbols $H^P$ and $H^Q$ always appear in the combination
\beq
H^P(\baz,z) + H^Q(\baz,z) = 2 H^W(\baz,z) (1 + O(\hbar^2)).
\eeq
The differential equation for $G_{ud}$ is
\beq
-i {\dot G}_{ud} = B^W - 2 A^W G_{ud} + \baB^W G_{ud}^2,
\eeq
where we have added a superscript W to show that the symbol for the Hamiltonian that is involved is $H^W$.
This differential equation is the same as before with the same initial conditions. Hence,
\beq
G_{ud}(t) = \frac{1}{\baB^W(t)} \Bigl(A^W(t) + i \frac{\dot v}{v} \Bigr).
\eeq

Finally, we need the product of all the determinants $\det G_k$. Keeping in mind that the determinant from
every other step is unity, we have
\bea
\ln K_{\rm red}
 &=& -\hf \sum_{k=1}^M \ln(\det G_k) \nnu\\
 &=& -\hf \int_0^T
         \Bigl[ i \frac{\ptl^2 H^P}{\ptl\baz\ptl z} 
                   - i \frac{\ptl^2 H^W}{\ptl z^2} G_{ud}(t) \Bigr]\, dt \nnu\\
 &=& -\frac{i}{2} \int_0^T \Bigl[ A^P(t) - \Bigl(A^W(t) + i \frac{\dot v}{v} \Bigr) \Bigr]\,dt.
\eea
Now, $A^P - A^W = O(\hbar)$, which may be neglected since the term we are discussing is already the first correction
in powers of $\hbar$. Hence $\ln K_{\rm red} = -\hf \ln v(T)$, i.e.,
\beq
K_{\rm red} = \left( i\frac{\ptl^2 S^W}{\ptl \baz_f \ptl z_i} \right)^{1/2},
\eeq
which has no SK correction.
      
\acknowledgments
This work was supported in part by the NSF via grant numbers PHY-0854896 (F.~Li and A.~Garg),
DGE-0801685 (NSF-IGERT program) (C.~Braun), and DMR 13-06011 (M.~Stone).

\appendix
\section{Review of P, Q, and Weyl symbols for particles and spins}
\label{symbols}

\subsection{Mapping for particles}
\label{rev_part_pqw}

For a massive particle in one spatial dimension, the Q and P symbols of the Hamiltonian $\ham$ are defined by
\bea
H^Q(\baz, z) &=& \frac{\mel{\baz}{\ham}{z}}{\olap{\baz}{z}}, \label{def_HQ_diag} \\
\ham &=& \int {d^2z \over \pi} e^{-\baz z} H^P(\baz, z) \kb{z}{\baz}. \label{def_HP}
\eea
(Analogous definitions apply to other operators.)
In \eno{def_HP} $d^2z$ is shorthand for $dx\,dy$, with $x$ and $y$ being the real and imaginary parts of $z$.
The reason for the $e^{-\baz z}$ factor inside the integral for this equation and the $\olap{\baz}{z}$
denominator of the previous one is that as defined in \eno{def_cs_part}, the states $\ket{z}$ and
$\bra{\baz}$ are not normalized; instead
\beq
\olap{\baz}{z'} = e^{\baz z'}.
\eeq
The resolution of unity therefore takes the form
\beq
1 = \int {d^2z\over \pi} e^{-\baz z} \kb{z}{\baz},
\eeq
with the same extra $e^{-\baz z}$ factor. The advantage of using unnormalized states is that off-diagonal matrix
elements such as $\mel{\baz}{\ham}{z'}$ can be obtained from the diagonal one, $\mel{\baz}{\ham}{z}$, by appealing
to analyticity. In practical terms this means that we merely replace $z$ with $z'$.

For the Weyl symbol, we follow Weyl himself~\cite{weyl}, and define
\beq
H^W(\baz, z) = \Tr \bigl(\ham \wey(\baz, z) \bigr), \label{op_to_weyl}
\eeq
where $\wey(\baz, z)$ is an operator-valued kernel given by
\beq
\wey(\baz, z) = \int \frac{d^2w}{\pi} e^{(w \adag - \baw a)} e^{-(w \baz - \baw z)}. \label{weyl_kernel}
\eeq
This definition puts the familiar symmetrization rule for Weyl ordering of operators on a broader footing and
can be shown to reduce to that for simple examples such as $a^2 (\adag)^2$.  The Weyl symbol for any other
operator is defined analogously. By letting $w \to -w$, $\baw \to -\baw$ in \eno{weyl_kernel}, we find that
$[\wey(\baz, z)]^{\dagger} = \wey(\baz, z)$. Hermiticity of $\ham$ then implies that $H^W(\baz, z)$ is real.

One possible inverse of the transform (\ref{op_to_weyl}) is given by
\beq
\ham = \int \frac{d^2z}{\pi} H^W(\baz, z) \wey(\baz, z). \label{weyl_to_op}
\eeq
It is straightforward to show this result by using the identities (themselves easily shown)
\bea
\Tr \wey(\baz, z) &=& 1, \\
\Tr \bigl(\wey(\baz, z) \wey (\baz', z') \bigr) &=& \pi \dta^{(2)}(z-z'),
\eea
where by $\dta^{(2)}(z)$ we mean $\dta({\rm Re\,} z) \dta({\rm Im\,} z)$. In fact, the inverse
(\ref{weyl_to_op}) is unique, as may be shown by taking the trace in \eno{op_to_weyl} in the
complete set of position states. The matrix elements of the kernel $\wey$ are not difficult to find,
and the trace takes on the form of an ordinary Fourier integral, which may be inverted to obtain
an expression for the position-space matrix elements of $\ham$ in terms of the function $H^W$. This
expression is easily seen to be identical to the one implied by \eno{weyl_to_op}. Thus \eno{weyl_to_op}
not only implies but is also implied by \eno{op_to_weyl}.

Next, we recapitulate the relationship between the Q, P, and Weyl symbols~\cite{moy}.
Consider the matrix element
\beq
\mel{\baz_2}{\ham}{z_1},
\eeq
which is nothing but $\olap{\baz_2}{z_1} H^Q(\baz_2, z_1)$. (See \eno{def_HQ}.) Writing $\ham$ in
terms of $H^W(\baz, z)$, and $\olap{\baz_2}{z_1} = e^{\baz_2 z_1}$, we obtain
\beq
H^Q(\baz_2, z_1) = e^{-\baz_2 z_1} \int \frac{d^2z}{\pi} H^W(\baz, z)
                       \mel{\baz_2}{\wey(\baz, z)}{z_1}. \label{weyl_to_Q}
\eeq
Using \eno{weyl_kernel}, we get
\bea
\mel{\baz_2}{\wey(\baz, z)}{z_1} 
   &=& \int \frac{d^2w}{\pi} e^{-(w\baz - \baw z)}
                       \exp\bigl(\baz_2 w - \baw z_1 - \tshf \baw w + \baz_2 z_1\bigr) \nnu \\
   &=& \int \frac{d^2w}{\pi} 
            \exp\bigl(-\tshf \baw w + w(\baz_2 - \baz) - w (z_1 - z) + \baz_2 z_1 \bigr) \nnu\\
   &=& 2 \exp\bigl(-2(\baz_2 - \baz)(z_1 - z) + \baz_2 z_1 \bigr).
       \label{mel_weyl_ker}
\eea
We now feed this result into \eno{weyl_to_Q} while at the same time defining
\beq
\eta = z - z_1, \quad \bet = \baz - \baz_2. \label{shifts}
\eeq
We thus get
\beq
H^Q(\baz_2, z_1) = 2\int \frac{d^2\eta}{\pi} e^{-2\bet\eta} H^W(\baz_2+\bet, z_1 + \eta).
\eeq
If we now Taylor expand $H^W$ in powers of $\eta$ and $\bet$, it is easy to perform the resulting Gaussian
integrals. Retaining the first nonzero correction, we get
\beq
H^Q(\baz_2, z_1) = H^W(\baz_2, z_1) + \hf \frac{\ptl^2}{\ptl\baz_2 \ptl z_1} H^W(\baz_2, z_1) + \cdots.
   \label{HW_to_HQ}
\eeq
The second term in this expansion is in fact of order $\hbar$ relative to the first. One can see this point by
writing the quantities $z$ and $\baz$ in terms of dimensionful position and momentum variables, and noting that
$z$ and $\baz$ both contain a factor of $\hbar^{-1/2}$. By transposing this term to the left hand side, and using
the same equation recursively, we find that
\beq
H^W(\baz_2, z_1) = H^Q(\baz_2, z_1) - \hf \frac{\ptl^2}{\ptl\baz_2 \ptl z_1} H^Q(\baz_2, z_1) + \cdots.
   \label{HQ_to_HW}
\eeq

To relate $H^P$ and $H^W$, we substitute \eno{def_HP} in \eno{op_to_weyl}, and obtain
\bea
H^W(\baz_2, z_1)
   &=& \Tr \bigl(\ham \wey(\baz_2, z_1) \bigr) \nnu\\
   &=& \int \frac{d^2z}{\pi} e^{-\baz z} H^P(\baz, z)
          \Tr \bigl( \kb{z}{\baz} \wey(\baz_2, z_1) \bigr) \label{weyl_to_P}
\eea
Now,
\bea
\Tr \bigl( \kb{z}{\baz} \wey(\baz_2, z_1) \bigr)
    &=& \mel{\baz}{\wey(\baz_2, z_1)}{z} \nnu\\
    &=& 2 \exp\bigl(-2(\baz - \baz_2)(z - z_1) + \baz z \bigr),
\eea
where the last result is obtained from \eno{mel_weyl_ker} with the exchange
$(\baz_2, z_1) \tofro (\baz, z)$.
Feeding it into \eno{weyl_to_P} along with the definitions (\ref{shifts}), we get
\beq
H^W(\baz_2, z_1) = 2\int \frac{d^2\eta}{\pi} e^{-2\bet\eta} H^P(\baz_2+\bet, z_1 + \eta).
\eeq
We now Taylor expand $H^W$ in powers of $\eta$ and $\bet$ just as done above, and 
integrate over $\eta$ and $\bet$. Again retaining only the first nonzero correction, we get
\beq
H^W(\baz_2, z_1) = H^P(\baz_2, z_1) + \hf \frac{\ptl^2}{\ptl\baz_2 \ptl z_1} H^P(\baz_2, z_1) + \cdots.
   \label{HP_to_HW}
\eeq

For the particle case, we can achieve a more general correspondence between operators and phase-space functions
by extending the definition of the Weyl kernel to~\cite{ber80}
\beq
\wey^{(\al)}(\baz, z)
   = \int \frac{d^2w}{\pi}
        e^{(w \adag - \baw a)}
          e^{-(w \baz - \baw z)} e^{\al \baz z/2}, \label{alfa_kernel}
\eeq
where $-1\le \al \le 1$. The mapping is then given by
\bea
H^{(\al)}(\baz, z) &=& \Tr \bigl(\ham \wey^{(\al)}(\baz, z) \bigr), \label{op_to_alfa} \\
\ham &=& \int \frac{d^2z}{\pi} H^{(\al)}(\baz, z) \wey^{(-\al)}(\baz, z). \label{alfa_to_op}
\eea
The cases of Q, P, and Weyl mappings correspond to $\al$ = $1$, $-1$, and $0$ respectively. We shall not
employ this general definition, but shall work with the P and Q mappings in the form given earlier.

\subsection{Mapping for spin}
\label{rev_spin_pqw}

Let us first discuss what we mean by the Hamiltonian of a spin system. For a particle of spin $j$, the most
general Hamiltonian (or any other operator) can be written as~\cite{herm_ham}
\beq
\ham = \sum_{\ell = 0}^{2j} \sum_{m=-\ell}^{\ell} c_{\ell m} \Ylm(\bJ),
\eeq
where $c_{\ell m}$ are arbitrary c-number coefficients, and
$\Ylm$ are spherical harmonic tensor operators defined via the operator analogue of the Herglotz generating
function for spherical harmonics. The classical phase space can be taken as a sphere of fixed radius
(which may be taken as 1, or $j$, or $j+\tshf$, whichever is most convenient). As discussed in \Sno{spinW}, the
variables $z$ and $\baz$ that we employed in \Sno{spin_prop} are stereographic coordinates for this sphere.
It is somewhat easier at first, however, to parametrize a point on this sphere by its direction $\nhat$,
so that functions on phase space are functions of $\nhat$. The Q, Weyl, and P symbols for the Hamiltonian are
\beq
H^{Q, W, P}(\nhat)
    = \sum_{\ell = 0}^{2j} \sum_{m=-\ell}^{\ell} c_{\ell m} \Phi_{\ell m}^{Q, W, P}(\nhat),
\eeq
where $\Phi_{\ell m}^{Q, W, P}(\nhat)$ are the corresponding symbols for $\Ylm(\bJ)$. We gave complete
expressions for these in Ref.~\olc{lbg13}, but here we only need the asymptotic forms as $j \to \infty$.
Recalling the definition
\beq
\jtil = j + \tshf,
\eeq
we have
\bea
\wlm(\nhat)
  &\apx& \jtil^{\ell} \bigl(1 + O(\jtil^{-2}) \bigr) \ylm(\nhat), \\
\Phi^{Q, P}(\nhat)
  &\apx& \jtil^{\ell} \Bigl(1 \mp \frac{\ell(\ell +1)}{4\jtil} +O(\jtil^{-2}) \Bigr) \ylm(\nhat).
\eea
We now observe that
$
\ell(\ell +1) \ylm(\nhat) = \am^2 \ylm(\nhat),
$
where
$
\am = - i (\nhat \times \grad_{\nhat})
$
is the angular momentum operator (on phase space, and {\it not\/} the quantum mechanical Hilbert space).
Hence, we may write
\beq
\Phi_{\ell m}^{Q, P}
  = \Bigl(1 \mp \frac{\am^2}{4\jtil} +O(\jtil^{-2}) \Bigr) \wlm.
\eeq
It follows that
\beq
H^{Q, P}
  = \Bigl(1 \mp \frac{\am^2}{4\jtil} +O(\jtil^{-2}) \Bigr) H^W,
    \label{HW_to_HPQ}
\eeq
a result which makes no reference to $\ylm(\nhat)$ and is therefore valid independent of the form in
which the Weyl symbol is given. It has a pleasing similarity to \eno{HW_to_HQ} etc.\ if we recall
that $-\am^2$ is (the angular part of) the Laplacian on the sphere.

\vskip20pt
\noindent $^{\dagger}$Present address: Online School for Girls, 7303 River Rd, Bethesda, MD 20817. 
\noindent $^{\dagger\dagger}$Present address: Department of Radiation Oncology, University of Florida,
            P.O. Box 100385, Gainesville, Florida 32610

%\newpage
%\voffset=1.0in
%\hoffset=0.5in
%
%\begin{figure}
%\includegraphics{mvt_direct.ps}
%%\centerline{\epsfig{figure=mvt_loglog.eps,height=2in}}
%\caption{\label{mvt_direct.ps}
%Same as \fno{m_vs_t_short.eps}, except that the rate equations are solved using $\kap_2 = 50$.
%} % close caption bracket
%\end{figure}
%

\begin{thebibliography}{99}
\bibitem{no_Q_in07}
The notation $H^Q$ was not employed in Refs.~\onlinecite{spg00,bra07a}, but it clearly makes sense to do so
now.
\bibitem{trace_pq}
We observe that a similar traciality rule holds in the particle case also, but there it is generally taken as
a consequence of the basic definition of the Weyl symbol rather than as a defining principle itself.
\bibitem{grad2}
This result may be found efficiently by using the fact that $\am^2$ is the negative of the Laplace-Beltrami
operator $\Dta$, and then using formulas from geometry: If $x^i$ are arbitrary coordinates in a curved space
and $g^{ik}$ is the contravariant metric tensor, then $\Dta = g^{ik} \ptl^2/\ptl x^i \ptl x^k$. Taking
$x^1 =z$ and $x^2 = \baz$, we first note that the squared length element on the unit sphere is
$ds^2 = 4(1+\baz z)^{-2} dz\,d\baz$, so that $g_{zz} = g_{\baz\baz} = 0$ and $g_{z\baz} = g_{\baz z} = 2(1+\baz z)^{-2}$.
Since the covariant and contravariant metrics are reciprocals, i.e., $g^{ik}g_{km} = \dta^i_m$, we have
$g^{zz} = g^{\baz\baz} = 0$, $g^{z\baz} = g^{\baz z} = (1+\baz z)^2/2$. The result then follows.
\bibitem{not_herm}
In saying that the matrix is ``symmetric," we are following physicists' usage, by which is meant that
$D_{ij} = D_{ji}$, not that $D_{ij} = D^*_{ji}$.
\end{thebibliography}
\end{document}